\documentclass[a4paper]{article}
\usepackage{amssymb}
\usepackage{graphicx}
\usepackage{algorithm}
\usepackage{algorithmic}
\usepackage{diagbox}
\usepackage{color}
\usepackage{amsmath}
\usepackage{multirow}
\usepackage{array}
\usepackage{hyperref}
\usepackage{cite}

\usepackage{INTERSPEECH2020}

\title{Emotion Profile Refinery for Speech Emotion Classification}
\name{Shuiyang Mao, P. C. Ching, Tan Lee}
\address{Department of Electronic Engineering, The Chinese University of Hong Kong, Hong Kong}
\email{maoshuiyang@link.cuhk.edu.hk, pcching@ee.cuhk.edu.hk, tanlee@ee.cuhk.edu.hk}

\begin{document}

\maketitle
\begin{abstract}
Human emotions are inherently ambiguous and impure. When designing systems to anticipate human emotions based on speech, the lack of emotional purity must be considered. However, most of the current methods for speech emotion classification rest on the consensus, e.\,g., one single hard label for an utterance. This labeling principle imposes challenges for system performance considering emotional impurity. In this paper, we recommend the use of emotional profiles (EPs), which provides a time series of segment-level soft labels to capture the subtle blends of emotional cues present across a specific speech utterance. We further propose the emotion profile refinery (EPR), an iterative procedure to update EPs. The EPR method produces soft, dynamically-generated, multiple probabilistic class labels during successive stages of refinement, which results in significant improvements in the model accuracy. Experiments on three well-known emotion corpora show noticeable gain using the proposed method.
\end{abstract}

\section{Introduction}
Automatic detection of human emotion in natural expressions is non-trivial. This difficultly is in part due to emotional ambiguity and impurity \cite{mower2010framework}. However, conventional emotion classification systems rely on majority voting (i.\,e., one-hot hard label) from a set of annotators as the ground truth. This labeling principle imposes specific challenges on emotion classification tasks: 1) Incomplete Labeling: Human expressions involve a complex range of mixed emotional manifestations \cite{mower2009interpreting}. Emotion classification systems designed to output one emotion label per input speech utterance/segment may perform poorly if the expressions cannot be well captured by a single emotional label \cite{mower2010framework}. 2) Inter-category Dependency: Certain emotion classes are inherently ambiguous. For example, the emotion class of frustration has the potential to overlap with categories ranging from anger, to neutrality and to sadness \cite{mower2009interpreting, busso2008iemocap}.

Soft labeling approaches have been recently developed to characterize blended emotional expressions. For instance, Lotfian et al. devised an innovative probabilistic method for soft labeling of emotions \cite{lotfian2017formulating}. Ando et al. developed a \textit{deep neural network} (DNN)-based model trained with soft emotion labels as ground truth, to better characterize the emotional ambiguity \cite{ando2018soft}. Kim et al. proposed to use cross entropy to directly compare human and machine emotion label distributions based on soft labeling \cite{kim2018human}. 

While soft labeling provides better flexibility in characterizing the emotional impurity and ambiguity, in most of the existing work, the soft labels are assigned per utterance, which is termed static soft labeling. However, as is well known, emotions in natural human expressions do not follow a static mold. Instead, they vary temporally with speech \cite{mower2009interpreting, busso2007interrelation}. The static soft labeling thus fails to characterize the emotional fluctuation across the utterance. A natural solution to this problem is to perform segment-level soft labeling. As a first step toward this goal, this work adopts an emotion classification paradigm based on \textit{emotion profiles} (EPs), which is a time series of segment-level soft labels across an utterance, with each dimension representing a classifier-derived probability of a possible emotion component. 

EPs have been around within the community for a while. For instance, Mower et al. derived EPs using a set of binary \textit{support vector machine} (SVM) outputs \cite{mower2010framework, provost2012simplifying}. Han et al. utilized a DNN-based model trained with stacked raw acoustic features to obtain deep-learned EPs \cite{han2014speech}. Our previous work further extended EPs into an end-to-end approach using a \textit{deep convolutional neural networks} (DCNN) \cite{mao2019deep}. While these EPs based studies have achieved impressive performance and provided more interpretable representations than traditional systems, one major shortcoming remains: the lack of segment-level ground truth labels. To circumvent this problem, most of the previous studies assigned the utterance-level one-hot label, which we call \textit{pseudo one-hot label}, to all of the segments within the same utterance \cite{han2014speech, mao2019deep}, or trained the segment-level classifier with utterance-level dataset \cite{provost2012simplifying}. This may result in an inconsistency with the ground truth or impart a mismatch to the segment-level classifier.

To better train a segment-level classifier, we argue that several characteristics should apply to ideal segment-level labels: 1) Labels should be informative of the specific segment, meaning that they should not be identical for all the segments across a given utterance. Therefore, labels should be defined at the segment-level rather than merely inheriting the label of the whole utterance. 2) Determining an ideal label for each segment may require observing the entire data to establish intra- and inter-category relations, suggesting that labels should be collective across the whole dataset. To achieve this, we propose emotion profile refinery (EPR). This solution uses a neural network model and the data to dynamically update the segment-level labels during the successive stages of refinery, enabling to generate more informative and collective segment-level labels. 

Extensive experiments are conducted on three popular emotion corpora, namely, the CASIA corpus \cite{tao2008design}, the Emo-DB corpus \cite{burkhardt2005database} and the SAVEE database \cite{jackson2014surrey}. Experimental results show that the proposed method consistently improves the accuracy of models for speech emotion classification by a significant margin: the CASIA corpus from \(93.10\%\) to \(94.83\%\) (WA\&UA), the Emo-DB corpus from \(83.00\%\) to \(88.04\%\) (WA) and \(82.36\%\) to \(87.78\%\) (UA), and the SAVEE database from \(70.63\%\) to \(77.08\%\) (WA) and \(69.88\%\) to \(74.64\%\) (UA). Our contributions include: 1) proposing the EPR framework for speech emotion classification task, 2) achieving the state-of-the-art accuracy on the three emotion corpora, and 3) demonstrating the ability of a network to improve accuracy by training from labels generated by another network of the same architecture.

\section{Methods}
\begin{figure}[tp]
  \centering
  \includegraphics[width=0.8\linewidth]{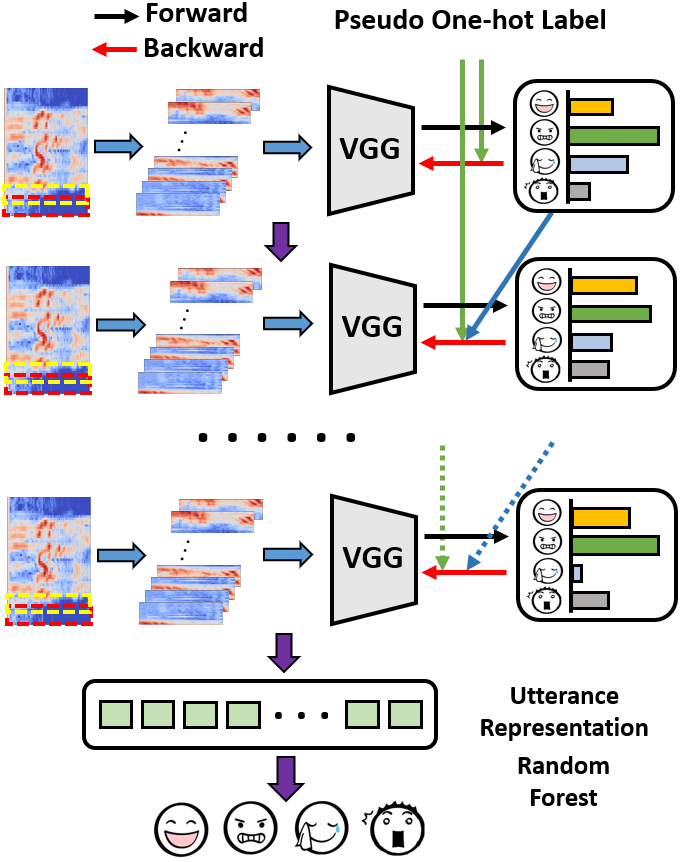}
  \caption{Illustration of the proposed method}
  \label{fig:framework}
\end{figure}
Figure~\ref{fig:framework} illustrates a schematic approach of the proposed method. It comprises a series of VGG \cite{simonyan2014very} networks trained to generate EPs from log-Mel filterbanks of individual segments. As the networks go through various stages of the refinery, the segment-level labels (and hence the EPs) are updated. The latest EPs are used for constructing utterance representations (i.\,e., extracting statistics across the EPs as in \cite{mao2019deep}). Finally, a \textit{random forest} (RF) is employed to assign the utterance-level labels. 

\subsection{Emotion profiles (EPs)}
Emotion profiles (EPs) were investigated and demonstrated to be useful for emotion classification tasks in \cite{mower2010framework, provost2012simplifying, han2014speech, mao2019deep, mao2018effective, shangguan2015emoshapelets, kim2013emotion}. Typically, EPs are time series of classifier-derived segment-level estimates of a set of the ``basic” emotions (e.\,g., angry, happy, neutral, sad), with each EP component representing the probability of the corresponding emotion category.

\subsubsection{Generating EPs}
We generate the EPs using a VGG model trained on the 64-bin log Mel filterbanks of individual segments. The log Mel filterbanks are computed by \textit{short-time Fourier transform} (STFT) with a window length of 25 ms, hop length of 10 ms, and FFT length of 512. Subsequently, 64-bin log Mel filterbank features are derived from each short-time frame, and the frame-level features are combined to form a time-frequency matrix representation of the segment. The trained VGG model aims to predict a probability distribution \(\textbf{\textit{P}}_i\) for the \(i^{\textnormal{th}}\) segment in Utterance \(\textbf{\textit{U}}\):
\begin{align}
  \textbf{\textit{P}}_i = [p_i(e_1),\ p_i(e_2),\ \cdots,\ p_i(e_K)]^T \in \mathbb{R}^{K\times 1}
  \label{eq1}
\end{align}
where, \(e_{1},\ e_{2},\ \cdots,\ e_{K}\), represent the set of ``basic" emotions, and \(K\) denotes the number of possible emotions. The EP for Utterance \(\textbf{\textit{U}}\) can then be formed as a multi-dimensional signal:
\begin{align}
  \textbf{\textit{U}}_{EP} = [\textbf{\textit{P}}_1,\ \textbf{\textit{P}}_2,\ \cdots,\ \textbf{\textit{P}}_{N}] \in \mathbb{R}^{K\times N}
  \label{eq2}
\end{align}
where \(N\) is the number of segments in the utterance.

\subsection{Emotion profile refinery (EPR)}
Simply assigning the utterance-level emotion label to all of its segments as the ground truth may not be accurate. We address this problem by passing the dataset through multiple EPs refiners (i.\,e., a series of VGG networks). The first refinery network \(C_1\) is trained over the dataset, where each training segment is assigned the pseudo one-hot hard label that inherited from its utterance. The second refinery network \(C_2\) is trained over the same dataset but uses soft labels generated by \(C_1\) (maybe combined with the original pseudo one-hot hard labels to mitigate an overfitting problem caused by the refinery process, which will be discussed in Section 4). Once \(C_2\) is trained, we can similarly use the updated EPs to train a subsequent network \(C_3\), and so on. The latest EPs are used as the ground truth EPs to construct the utterance representations for further classification.

\subsubsection{Loss}
We train the first refinery VGG network \(C_1\) using the cross-entropy loss against the pseudo one-hot labels. We train each of the subsequent refinery networks \(C_t\) for \(t > 1\) by minimizing the KL-divergence between its output and the soft label (maybe combined with the original pseudo one-hot hard label) generated by the previous refinery network \(C_{t-1}\). Letting \(p^{t}(e_k)\) be the probability assigned to class \(e_k\) in the output of model \(C_t\), our loss function for training model \(C_t\) is:
\begin{equation} \label{eq:kl-loss}
\begin{aligned}
\mathcal{L}_t &= -\sum_{k}p^{t-1}(e_k)\;\textnormal{log}\frac{p^{t}(e_k)}{p^{t-1}(e_k)}\\
&= -\sum_{k}p^{t-1}(e_k)\;\textnormal{log}p^{t}(e_k)+\sum_{k}p^{t-1}(e_k)\;\textnormal{log}p^{t-1}(e_k)
\end{aligned}
\end{equation}
The second term is constant with respect to \(C_t\). We can remove it and instead minimize the cross-entropy loss:
\begin{equation} \label{eq:ce-loss}
\begin{aligned}
\hat{\mathcal{L}}_t = -\sum_{k}p^{t-1}(e_k)\;\textnormal{log}p^{t}(e_k)
\end{aligned}
\end{equation}

\section{Emotion Corpora}
Three different emotion corpora are used to evaluate the validity and universality of our method, namely, a Chinese emotion corpus (CASIA) \cite{tao2008design}, a German emotion corpus (Emo-DB) \cite{burkhardt2005database} and an English emotional database (SAVEE) \cite{jackson2014surrey}, which are summarized in Table~\ref{tab:data}. All of the emotion categories are selected for each of the three stated emotion corpora, respectively.

Specifically, the CASIA corpus \cite{tao2008design} contains \(9,600\) utterances that are simulated by four subjects (two males and two females) in six different emotional states, i.\,e., angry, fear, happy, neutral, sad, and surprise. In our experiments, we only use \(7,200\) utterances that correspond to \(300\) linguistically neutral sentences with the same statements.

The Berlin Emo-DB German corpus (Emo-DB) \cite{burkhardt2005database} was collected by the Institute of Communication Science at the Technical University of Berlin. Ten professional actors (five males and five females) each produced ten utterances in German to simulate seven different emotions. The number of spoken utterances for these seven emotions is not equally distributed: \(126\) anger, \(81\) boredom, \(47\) disgust, \(69\) fear, \(71\) joy, \(79\) neutral, and \(62\) sadness.

The Surrey audio-visual expressed emotion database (SAVEE) \cite{jackson2014surrey} consists of recordings from four male actors in seven different emotions: anger, disgust, fear, happy, sad, surprise, and neutral. Each speaker produced \(120\) utterances. The sentences were chosen from the standard TIMIT corpus and phonetically-balanced for each emotion. 

\begin{table}[thbp]
\caption{Overview of the selected emotion corpora. (\#Utterances: number of utterances used, \#Subjects: number of subjects, and \#Emotions: number of emotions involved.)}
\centering
\resizebox{1.0 \linewidth}{!}{%
\renewcommand\arraystretch{1.5}
\huge
\begin{tabular}{|p{3.6cm}<{\centering}|p{3.8cm}<{\centering}|p{4.0cm}<{\centering}|p{4.2cm}<{\centering}|p{3.6cm}<{\centering}|}
\hline
Corpora  & Language & \#Utterances & \#Subjects  & \#Emotions \\
\hline
CASIA  & Chinese & 7,200 & 4 (2 female) & 6 \\
Emo-DB  & German & 535  & 10 (5 female) & 7 \\
SAVEE  & English & 480 & 4 (0 female) & 7 \\
\hline
\end{tabular}%
}
\label{tab:data}
\end{table}

\section{Experiments}
We evaluate the proposed method on the three mentioned emotion corpora. We first explore the effect of EPR without combining the original pseudo one-hot hard label, which we call \textit{standard EPR} (sEPR). We then present some ablation studies and analyses to investigate the source of the improvements using the sEPR method. Finally, the original pseudo one-hot hard label is combined with the soft label generated by an iterative EPR process, which we call \textit{pseudo one-hot hard label assisted EPR} (pEPR). The pEPR method achieves the best results.

\subsection{Setup}
The size of each speech segment is set to 32 frames, i.e., the total length of a segment is 10 ms \(\times\) 32 + (25 - 10) ms = 335 ms. For the CASIA corpus, the segment hop length is set to 30 ms, whilst it is set to 10 ms for the Emo-DB corpus and the SAVEE database. In this way, we collected 418,722 segments for the CASIA corpus, 131,053 segments for the Emo-DB corpus, and 51,027 segments for the SAVEE database, to train the VGG network, respectively.

For the VGG network, the architecture of the convolutional layers is based on the configurations (i.\,e., configuration E) in the original paper \cite{simonyan2014very}. A tweak is made to the number of units in the last softmax layer in order to make it suitable for our tasks. In the training stage, ADAM \cite{kingma2014adam} optimizer with default setting in Tensorflow \cite{abadi2016tensorflow} was used, with an initial learning rate of \(0.001\) and an exponential decay scheme with a rate of \(0.8\) every \(2\) epochs. The batch size was set to \(128\). Early stopping with patience of \(3\) epochs was utilized to mitigate an overfitting problem. Maximum number of epochs was set to \(20\).

The EPs were generated using ten-fold cross-validation. A \textit{random forest} (RF) with default setting in Scikit-learn \cite{pedregosa2011scikit} was then employed to make the utterance-level decision, where another ten-fold cross-validation was performed. The results were presented in terms of unweighted accuracy (UA) and weighted accuracy (WA), respectively. It is worth noting that the UA and WA are the same for the CASIA corpus as the CASIA corpus is (perfectly) balanced concerning the emotion category.

\subsection{Standard EPR (sEPR)}
\begin{figure}[htp]
  \centering
  \includegraphics[width=1\linewidth]{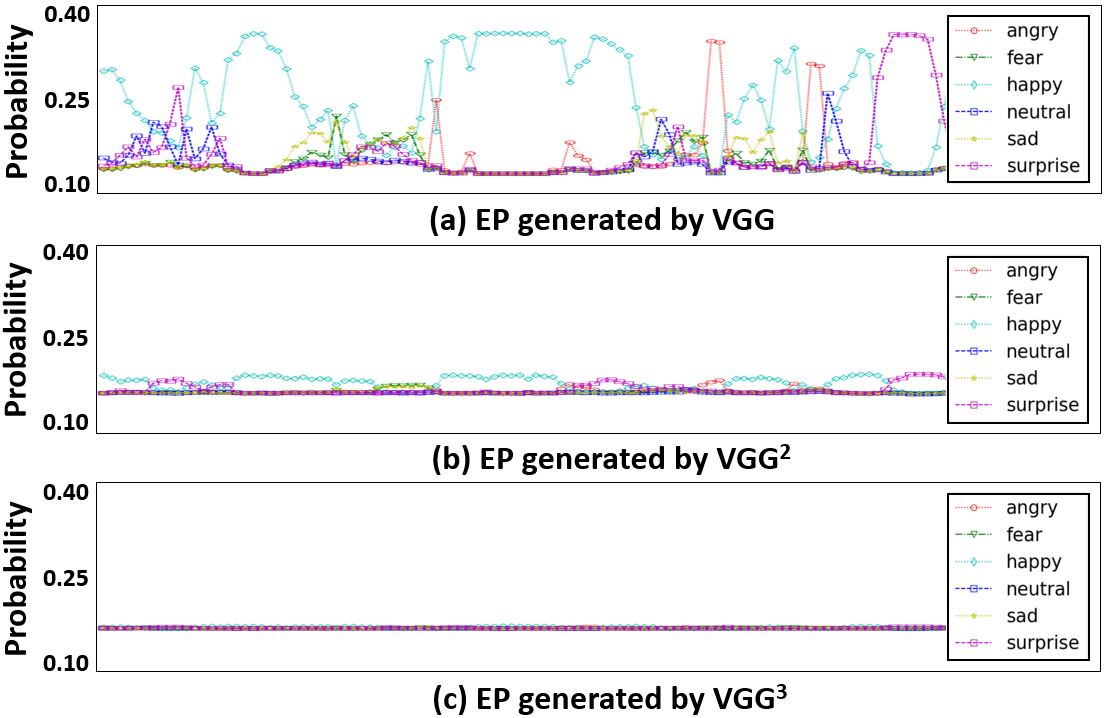}
  \caption{An example of EP evolution for the audio file ``Happy\_liuchanhg\_440.wav" from the CASIA corpus. The sEPR method was applied.}
  \label{fig:EPs_no_onehot}
\end{figure}
We first investigated the effect of sEPR. Table~\ref{tab:soft_no_hard} shows the experimental results on the three mentioned emotion corpora. Each row represents a randomly-initialized instance of VGG network trained with labels refined by the network directly one row above it in the table. As can be observed: 1) All VGG networks achieved the best performance after one single round of sEPR process, after which performance diminished significantly. 2) The performance gain was only minor. To explain these observations, we looked into the EPs generated during each sEPR iteration. Figure~\ref{fig:EPs_no_onehot} shows an example of EP evolution during two successive stages of refinement for the audio file ``Happy\_liuchanhg\_440.wav" from the CASIA corpus. It is obvious that the sEPR method tends to flatten and collapse the EPs iteratively, and each output dimension of \(\textnormal{VGG}^2\) is close to \(0.16\), i.\,e., the value obtained by a random guess for the CASIA corpus. We argue that this is because the model tends to minimize the cross-entropy progressively, and the refined EPs contain information that it has memorized from the previous round of training examples. Therefore, a severe overfitting problem happened. We further argue that there is a trade-off between the minimization of segment-level cross-entropy and the maximization of utterance-level accuracy. To 
address this problem, the pEPR method was proposed and experimented. This is discussed further in Section 4.4.
\begin{table}[htbp]
\renewcommand\arraystretch{1}
\caption{Results using the sEPR method on the three stated emotion corpora. Each model is trained using labels refined by the model right above it. That is, \(\textnormal{VGG}^2\) is trained by the labels refined by \(\textnormal{VGG}\), and so on. The first row networks are trained using the original pseudo one-hot hard labels.}
\resizebox{1. \linewidth}{!}{%
\begin{tabular}{l|c|c|c|c|c|c}
\toprule
\multicolumn{1}{c|}{} & \multicolumn{2}{c|}{CASIA} & \multicolumn{2}{c|}{Emo-DB} &\multicolumn{2}{c}{SAVEE} \\
\midrule
\bf Model & WA & UA & WA & UA & WA & UA\\
\midrule
VGG & $93.10$ & $93.10$ & $83.00$ & $82.36$ & $70.63$ & $69.88$\\
\(\textnormal{VGG}^2\) & $\bf93.67$ & $\bf93.67$ & $\bf83.74$  & $\bf83.96$ & $\bf71.88$ & $\bf70.64$\\
\(\textnormal{VGG}^3\) & $90.07$ & $90.07$ & $69.91$ & $67.92$ & $26.04$ & $21.07$\\
\bottomrule
\end{tabular}%
}
\label{tab:soft_no_hard}
\end{table}
\begin{figure*}[thbp]
  \centering
  \includegraphics[width=\linewidth]{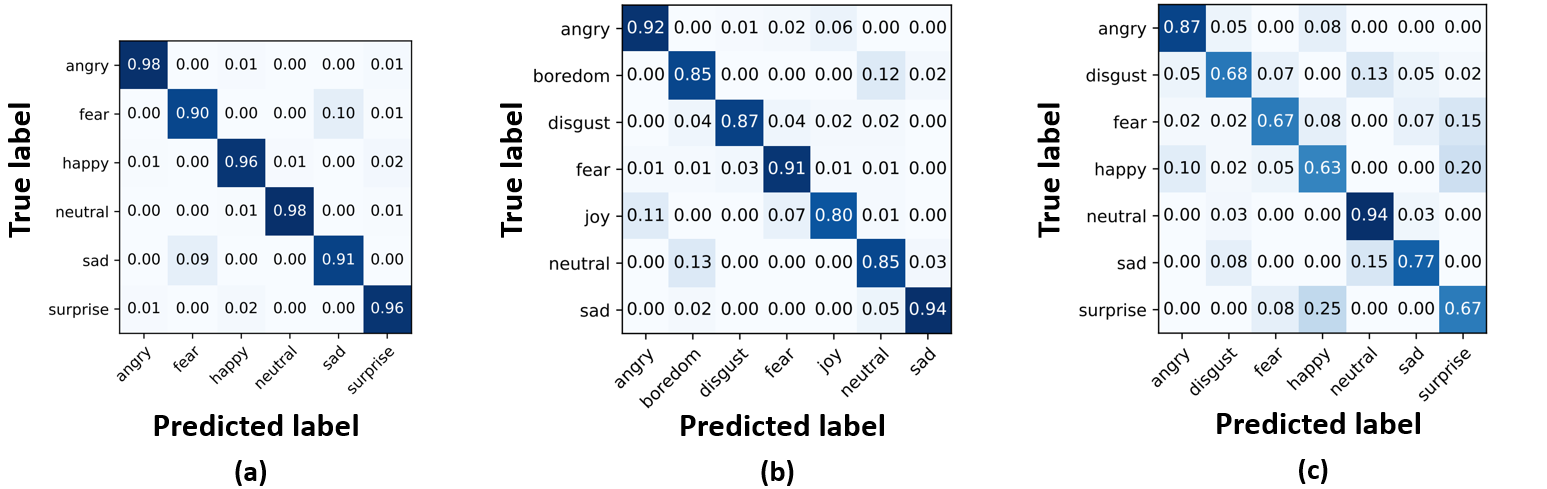}
  \caption{Confusion matrices obtained using the pEPR method on (a) the CASIA corpus; (b) the Emo-DB corpus; (c) the SAVEE database.}
  \label{fig:confusionM}
\end{figure*}

\subsection{Dynamic labels vs. soft labels}
In the very beginning, we posit that the benefits of using sEPR are twofold: 1) Each segment is dynamically re-labeled with a more accurate label, and 2) the introduction of soft labeling. To assess the improvement from dynamic labeling alone, we performed label refinement with hard dynamic labels. Specifically, we passed each segment to the VGG network, and the one-hot label was assigned by choosing the most-likely category from the network output. To observe the improvement from soft labeling alone, we investigated the soft static labels. To compute the soft static label for a given segment, we passed all segments within the same utterance to the VGG network, and the soft static label was computed by averaging the network outputs across the utterance. Table~\ref{tab:soft_n_hard} shows the results. As can be seen, the hard dynamic labeling consistently improved the accuracy of the network for the three emotion corpora, while it was not the case for the soft static labeling. However, when they were combined we observed an additional improvement, suggesting that they address different issues with labels in the dataset.
\begin{figure}[tp]
  \centering
  \includegraphics[width=1\linewidth]{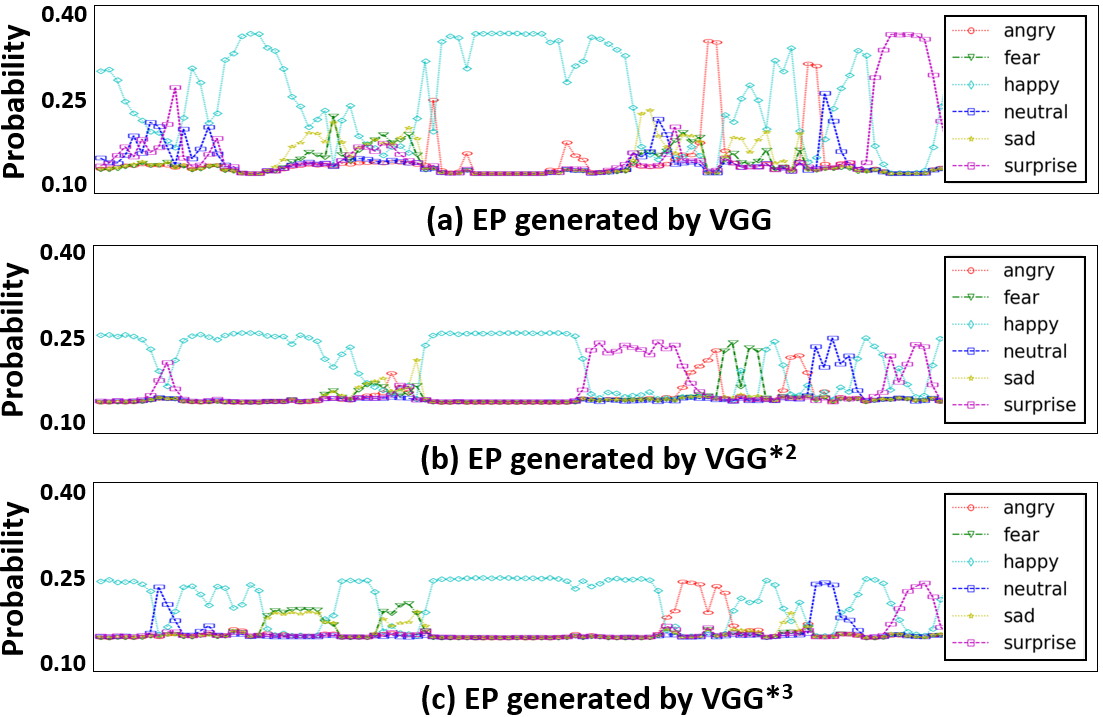}
  \caption{An example of EP evolution for the audio file ``Happy\_liuchanhg\_440.wav" from the CASIA corpus. The pEPR method was applied.}
  \label{fig:EPs}
\end{figure}
\begin{table}[htbp]
\renewcommand\arraystretch{1}
\caption{Comparison of experimental results for  hard dynamic labels and soft static labels.}
\resizebox{1. \linewidth}{!}{%
\begin{tabular}{l|c|c|c|c|c|c}
\toprule
\multicolumn{1}{c|}{} & \multicolumn{2}{c|}{CASIA} & \multicolumn{2}{c|}{Emo-DB} &\multicolumn{2}{c}{SAVEE} \\
\midrule
\bf Model & WA & UA & WA & UA & WA & UA\\
\midrule
No Refinery & $93.10$ & $93.10$ & $83.00$ & $82.36$ & $70.63$ & $69.88$\\
Soft Static & $91.36$  & $91.36$  & $79.64$ & $78.77$ & $67.71$ & $66.48$ \\
Hard Dynamic & $93.21$ & $93.21$ & $83.18$ & $83.13$ & $71.04$ & $70.00$\\
Soft Dynamic &  $\bf93.67$ & $\bf93.67$ & $\bf83.74$ & $\bf83.96$ & $\bf71.88$ & $\bf70.64$\\
\bottomrule
\end{tabular}%
}
\label{tab:soft_n_hard}
\end{table}

~\\[-1cm]
\subsection{Pseudo one-hot hard label assisted EPR (pEPR)}

In this section, we aimed at mitigating the overfitting problem reported in Section 4.2. We handled this issue by combining the generated soft labels with the original pseudo one-hot hard labels. Specifically, the network output (e.\,g., [$0.6$, $0.1$, $0.1$, $0.2$]) of a certain segment and its original pseudo one-hot hard label (e.\,g., [$1$, $0$, $0$, $0$]) were added and normalized (i.\,e., [$0.8$, $0.05$, $0.05$, $0.1$]), which was then used as the refined label to train the next network. The intuition of this operation is only natural. Since there exists a trade-off between the minimization of the segment-level cross-entropy and the optimization of the utterance-level performance (refer to Section 4.2), we conjecture that the combination of the original pseudo one-hot hard labels might offer an advantage in regularizing the segment-level network training and adding a strong bias towards utterance-level accuracy. Figure~\ref{fig:EPs} shows an example of EP evolution generated using the pEPR method for the same audio file as in Section 4.2. It can be observed that the serve EPs flattening and collapse encountered using sEPR method (see Figure~\ref{fig:EPs_no_onehot}) disappeared. Table~\ref{tab:soft_n_hard_onehot} shows the results. A significant improvement can be observed compared to the sEPR method, which corroborated our previous conjecture. Figure~\ref{fig:confusionM} shows the corresponding confusion matrices obtained using the pEPR method on the three mentioned emotion corpora, respectively.

\begin{table}[htbp]
\renewcommand\arraystretch{1}
\caption{Results using the pEPR method on the three stated emotion corpora.}
\resizebox{1. \linewidth}{!}{%
\begin{tabular}{l|c|c|c|c|c|c}
\toprule
\multicolumn{1}{c|}{} & \multicolumn{2}{c|}{CASIA} & \multicolumn{2}{c|}{Emo-DB} &\multicolumn{2}{c}{SAVEE} \\
\midrule
\bf Model & WA & UA & WA & UA & WA & UA\\
\midrule
VGG & $93.10$ & $93.10$ & $83.00$ & $82.36$ & $70.63$ & $69.88$\\
\(\textnormal{VGG}^{\ast2}\) & $\bf94.83$ & $\bf94.83$ & $87.10$ & $86.78$ & $73.96$ & $71.67$ \\
\(\textnormal{VGG}^{\ast3}\) & $94.54$ & $94.54$ & $86.92$ & $86.42$ & $76.67$ & $74.33$ \\
\(\textnormal{VGG}^{\ast4}\) & $94.60$ & $94.60$ & $85.23$ & $85.07$ & $\bf77.08$ & $\bf74.64$ \\
\(\textnormal{VGG}^{\ast5}\) & $94.24$ & $94.24$ & $\bf88.04$ & $\bf87.78$ & $74.58$ & $73.10$ \\
\bottomrule
\end{tabular}%
}
\label{tab:soft_n_hard_onehot}
\end{table}

\section{Conclusions}
In this paper, we addressed the problem of emotional impurity encountered in speech emotion classification task using emotion profile refinery (EPR). This method allows us to dynamically label the speech segments with soft targets, which characterizes the probability distributions of the underlying mixture of emotions at segment level. Two EPR method, namely, the standard EPR (sEPR) and the pseudo one-hot hard label assisted EPR (pEPR), were proposed and investigated, and the latter significantly outperformed the former. We achieved the state-of-the-art results on three well-known emotion corpora, respectively.

\newpage
\bibliographystyle{IEEEtran}
\bibliography{mybib}

\end{document}